\documentstyle[epsfig,aps]{revtex}
\def \Ogw {{\Omega_{\rm gw}}}
\def \hOgw {{h_{100}^2\,\Omega_{\rm gw}}}
\def \SNR {{\left(\frac{{\rm S}}{{\rm N}}\right)_m}}
\def \hatO {{\hat{\Omega}}}
\def \To   {T_{\rm obs}}
\newcommand{\be}{\begin{equation}}
\newcommand{\ee}{\end{equation}}
\newcommand{\ba}{\begin{eqnarray}}
\newcommand{\ea}{\end{eqnarray}}
\newcommand{\m}{\langle}
\newcommand{\M}{\rangle}
\def \hrms {h_{\rm rms}}
\def \hc {h_{\rm c}}
\def\ltsima{$\; \buildrel < \over \sim \;$}
\def\simlt{\lower.5ex\hbox{\ltsima}}
\def\gtsima{$\; \buildrel > \over \sim \;$}
\def\simgt{\lower.5ex\hbox{\gtsima}}
\font\tenbb=msbm10
\font\sevenbb=msbm7
\font\fivebb=msbm5
\newfam\bbfam
\textfont\bbfam=\tenbb \scriptfont\bbfam=\sevenbb
\scriptscriptfont\bbfam=\fivebb

\begin{document}
\draft

\twocolumn[\hsize\textwidth\columnwidth\hsize\csname
@twocolumnfalse\endcsname
\preprint{............}
\title{Studying the anisotropy of the gravitational wave stochastic background with LISA}
\author{Carlo Ungarelli$^{1)}$ and Alberto Vecchio$^{2)}$}
\address{$^{1)}$ School of Computer 
Science and Mathematics, University of Portsmouth, Mercantile House, 
Hampshire Terrace, Portsmouth P01 2EG, UK  \\
$^{2)}$ School of Physics and Astronomy, The University of Birmingham,
Edgbaston, Birmingham B15 2TT, UK}

\maketitle
\begin{abstract}

A plethora of gravitational wave stochastic backgrounds populate the sensitivity window of
the Laser Interferometer Space Antenna.
We show that LISA can detect the anisotropy of the background corresponding 
to the multipole moments of order $l=2$ and $4$. The signal-to-noise ratio generated by galactic
white dwarf binary systems could be as high as $\approx 60$ for 3 yrs of integration, and
LISA could provide valuable information on the spatial 
distribution of a variety of galactic sources. We also show that the cross-correlation of the
data sets from two interferometers could marginally lead to meaningful upper-limits on the 
degree of isotropy of the primordial gravitational wave background.  

\end{abstract}
\pacs{PACS number(s): 04.30. -w,04.80.Nn} \vskip1pc]

Gravitational wave (GW) stochastic backgrounds generated in the
early-Universe and produced by large populations of faint 
astrophysical sources are among the most interesting signals for GW 
laser interferometers. Their detection will provide a new arena for 
early-Universe cosmology and compact-object astronomy. 
The Laser Interferometer Space Antenna (LISA)~\cite{LISA_ppa}
represents the key instrument in the low-frequency
band of the spectrum ($\sim 10^{-5}\, \mbox{Hz} - 10^{-1}\, \mbox{Hz}$), 
complementary to Earth-based detectors, 
operating in the window $\sim 1\, \mbox{Hz} - 10\, \mbox{kHz}$. LISA 
could prove to be the most suitable device to search 
for, and study stochastic signals:
(i) Several cosmological models predict a primordial GW stochastic background 
with a flat energy density spectrum $\Ogw (f)\sim \mbox{const}$
(see~\cite{Maggiore00}, and references therein, for a thorough review): 
indeed the spectral density of the signal $S_h(f)$ generated at the detector 
output steeply increases at low frequencies as $S_h(f) \propto \Ogw (f)/f^3$; 
(ii) Stochastic backgrounds generated by the incoherent super-position of GWs
from a variety of short-period binary systems (including close white dwarf binaries,
W UMa binaries, and neutron star binaries) both galactic~\cite{HBW90,BH97,PP98,IKPP01} and 
extra-galactic~\cite{KP98,SFMP00} are {\it guaranteed} signals for LISA, 
with several components that dominate the instrumental noise in the frequency range  
$10^{-4}\,\mbox{Hz} \simlt f \simlt  3 \times 10^{-3}\,\mbox{Hz}$.

The key issue in searching for stochastic backgrounds, and gaining 
insights into their production mechanism and the underlying physics, 
is to identify unambiguously the GW signal in one single LISA data 
stream. In fact, a stochastic
background is a random process which is intrinsically indistinguishable from 
the detector noise. We have discussed this issue in~\cite{UV01}, 
in the context of {\it perfectly isotropic} signals, showing the advantage
of cross-correlation experiments involving two data sets with 
uncorrelated noise, and pointing out fundamental sensitivity limits. More recently it has been shown that, 
by estimating the LISA instrumental noise using the ``symmetrized Sagnac observable''~\cite{TAE00}
it is possible to detect an isotropic stochastic background 
with a sensitivity close to the one that can be achieved 
through cross-correlation~\cite{BH01}. 
However, {\it several stochastic backgrounds are expected to be anisotropic}. 
In particular, the background generated by galactic sources
carries a strong signature due both to the peripheral location of the solar 
system in the galaxy, and the spatial 
distribution of the sources. The primordial GW background, although 
expected to be intrinsically 
isotropic to a high degree, presents a clear dipole structure
due to the proper motion of the galaxy with respect to the
cosmological rest frame.

In this paper we show that the peculiar orbital configuration of LISA
-- the instrument baricenter follows an Heliocentric orbit with period
$T = 1$ yr, and the detector plane, tilted by $60^\circ$ with respect to
the Ecliptic, counter-rotates with the same 
period T~\cite{LISA_ppa} -- can be exploited to design a suitable data analysis scheme to
detect and study anisotropic stochastic signals using the Michelson
configuration. This possibility was earlier suggested in~\cite{GP97}.
Here we carry out a rigorous analysis from the point of view of 
data analysis, following~\cite{AO97}, and we apply our results to the relevant 
astrophysical and cosmological scenarios.
We provide estimates for the optimal signal-to-noise 
ratio (SNR) as a function of the signal parameters, and we show
that, by detecting an anisotropic background, it is indeed possible to gain key 
insights on the spatial distribution of short-period binary systems.
We also discuss the possibility of obtaining meaningful upper-limits on the 
degree of isotropy of the primordial GW background, and comment on  
the improvement of the sensitivity and information extraction that would come from 
the correlation of the data streams of two separated and suitably 
oriented instruments.

We start by briefly reviewing the fundamental data analysis concepts;
we assume the reader familiar with the literature
on the detection of stochastic backgrounds in GW 
observations~\cite{Flanagan93,AR99,AO97}.
The GW background amplitude at any time $t$ and position $\vec{x}$
can be written  in the transverse-traceless gauge as
\ba
h_{ab}(\vec{x},t) & = & \sum_{A=+,\times}
\,\int d\hat{\Omega} \int_{-\infty}^{\infty} df
e^{2\pi i(ft-\hat{\Omega}\cdot\vec{x}/c)} \nonumber\\
& & \quad\quad\quad\quad \times e^{A}_{ab}(\hat{\Omega}) h_A(\hat{\Omega},f)\,;
\label{hab}
\ea
$e^{A}_{ab}$ ($a,b=1,2,3$) is the polarization tensor of the wave,
$h_A(\hat{\Omega},f)$ are the two independent polarization amplitudes
($A=+,\times$), and $\hat{\Omega}$ is the unit vector on the two-sphere 
along the 
wave's propagation direction. Here we consider stochastic signals 
that are Gaussian, unpolarized but {\it not} isotropic: 
$\langle{h_A}(f,\hat{\Omega})\rangle = 0$ and $\langle{h_{A}}(f,\hat{\Omega})
{h^*_{A'}}(f',\hat{\Omega}')\rangle = \delta_{A\,A'}
\delta(f-f')\delta^2(\hat{\Omega}-\hat{\Omega}') H(|f|)P(\hat{\Omega})$.
$\m . \M$ stands for the ensemble average, and we have assumed 
that the frequency of the signal is uncorrelated with respect 
to the angular distribution. $H(|f|)$ and $P(\hat{\Omega})$ are related to 
the spectrum $\Ogw(f) \equiv d\rho_{\rm gw}(f)/\rho_{\rm c}d\ln f$ by~\cite{AO97}
\be
\Ogw(f) = \frac{8\pi^2}{3\,H^2_0}|f|^3 H(|f|) \int_S d \hat{\Omega}P(\hat{\Omega})\,.
\label{Pom}
\ee
The one-sided power spectral density of the GW background simply reads
$S_h(|f|) = H(|f|)/8\pi$. $P(\hat{\Omega})$ represents
the contribution to $\Ogw(f)$ from the different directions in the
sky $\hatO$; in the context of our analysis, 
it is useful to expand $P(\hat{\Omega})$ in terms of the multipole moments
$P_{lm}$ as $P(\hat{\Omega})=\sum_{l,m}P_{l\,m}Y_{l}^{m}(\hat{\Omega})$. 
The signal strain at the interferometer output is $h(t)=D^{ab}(t)\,h_{ab}
(\vec{x}(t),t)$, 
where $D_{ab}= [\hat{u}_a(t)\hat{u}_b(t)-\hat{v}_a(t)\hat{v}_b(t)]/2$
is the detector response tensor, and 
$\hat{u}_a$ and $\hat{v}_a$ are the unit vectors pointing, at each time, 
in the direction of the instrument arms; $\vec{x}(t)$
describes the trajectory of the interferometer centre of mass, and for
LISA the explicit time-dependent expressions can be found in~\cite{UV01}. 
The detector output $o(t) = h(t) + n(t)$ is affected by the
instrumental noise $n(t)$, that we assume to be Gaussian and stationary,
with one-sided noise spectral density $S_n(f)$.

The key idea to detect (and study the properties of) an anisotropic GW stochastic
background is to break up the data set(s) of length $\To$ into several data chunks 
of much smaller size $\tau \ll T < \To$, over which the detector location and
orientation is essentially constant, and construct the new signal
\be
S(t)=\int_{-\infty}^{+\infty}df\, \tilde{o}_j(f,t)\,\tilde{Q}(f)\,\tilde{o}_k^*(f,t)\,.
\label{St}
\ee
In Eq.~(\ref{St}) the indexes $j$ and $k$ label the instruments, and 
$\tilde{o}(f;t) = \int_{t-\tau/2}^{t+\tau/2}dt'e^{-2\pi ift'}o(t')$ is 
the Fourier transform of the data chunk of length $\tau$, centered around $t$. $S(t)$ is  
the correlation function suitably weighted by the optimal (real) filter function $\tilde{Q}(f)$ 
(to be determined in order to maximize the signal-to-noise ratio)~\cite{AO97}. One then 
looks for peaks in $S(t)$ at multiples of the 
LISA rotation frequency $1/T$: because $\tau \ll T$ the correlation
will vary as LISA changes orientation. To fix ideas, we can set $\tau \sim 10^5$ sec
and $\To \sim 10^8$ sec. 
If one considers two separated instruments, an additional time scale appears:
the light-travel-time between the instrument centre-of-mass $\xi$. For two detectors in
Heliocentric orbit at distance D, $\xi \simeq 500\, (D/{\rm AU})$ sec. As most of the correlation between
two detectors builds up for frequencies $f \simlt 1/\xi$, there is significant
signal for time-scales shorter than $\tau$. 
Since the LISA motion is periodic, one can decompose the mean value of $S(t)$ as a Fourier
series:
\be
\m S(t) \M = \sum_{-\infty}^{+\infty}e^{i\, 2\pi m t/T} \m S_m\M\,
\label{Sav}
\ee
where each harmonics $S_m$ reads
\be
S_m = \frac{1}{\To} \int_0^{\To}  e^{-i\, 2\pi m t/T} S(t)\, dt\,.
\label{Sm}
\ee
{\it The amplitude of $S_m$ represents the observable} whose 
signal-to-noise ratio is $({\rm S/N})_m = \mu_m/\sigma_m$, where $\mu_m = |\m S_m \M|$, and
$\sigma_m^2 = \m S_m\,S_{-m} \M - |\m S_m \M|^2$. It is clear that 
the noise contribution to the ensemble average $\m\M$, and, as a consequence, the SNR of
each harmonics, is different whether $o_j$ and $o_k$ share correlated noise or not.
We will therefore analyze the two cases in turn.

We start by considering the LISA mission. In this case $o_j = o_k = o$. Evaluating the 
expectation values $\m . \M$, and defining a suitable positive semi-definite 
scalar product~\cite{AO97}, which allows us to
derive the explicit expression of $\tilde{Q}(f)$, the SNR of the $m$-th 
harmonics becomes
\be
\SNR= \To^{1/2}\,\Theta_m^{(0)}\,
\left\{\int_{0}^{\infty}df{\cal J}^0(f)
\right\}^{1/2}\,,
\label{snrm1}
\ee
where
\be
{\cal J}^{(0)}(f)\equiv
\frac{4\, S_h(f)^2}{4\, \gamma_0^2\,S_h(f)^2+20\,\gamma_0\,S_h(f)\,S_n(f)+25\, S_n(f)^2}\,,
\label{J0}
\ee
and
\be
\Theta_m^{(0)}=\left|\sum_{l=|m|}^{\infty}P_{lm}\gamma^{(0)}_{lm}\right| \,.
\label{Theta}
\ee
$\gamma_0 = 3/4$ is the overlap reduction function~\cite{Flanagan93,AR99}
for two co-located and co-aligned detectors, and
$\gamma^{(0)}_{lm}=5/(16\pi^2)\,
\int_{0}^{2\pi}d\alpha\int d\hat{\Omega}\, Y_{lm}(\hat{\Omega})
\sum_A\,\left[D_{ab}(\alpha)e^{ab}_A(\hat{\Omega})\right]^2$
are the overlap reduction functions corresponding to the $l$-th multipole 
moment and the $m$-th harmonic. Notice that for co-located and co-aligned instruments
$\gamma_0$ and $\gamma^{(0)}_{lm}$ do not depend on the frequency. It is straightforward 
to show that $\gamma^{(0)}_{lm}= 0$ for $l$ odd and $l > 4$. {\it LISA 
is therefore sensitive only to quadrupole ($l = 2$) and octupole 
($l = 4$) anisotropy, and is completely "blind" to the whole 
set of odd multipole moments}. This property affects both the signal detection
and parameter estimation: LISA is not sensitive to the (usually strongest) dipole anisotropy,
and only two multipole moments can be exploited in order to extract information.

The function ${\cal J}^{(0)}(f)$ in~(\ref{snrm1}) takes a particularly simple form in the two physically 
relevant cases: (i) If $h(t) \gg n(t)$ ({\it e.g.} for 
backgrounds generated by galactic binaries for $f \simlt 3$ mHz~\cite{HBW90,BH97}) 
then ${\cal J}^{(0)}(f) \simeq 1/\gamma_0^2$; (ii) For $h(t) \ll n(t)$
(the likely situation for primordial backgrounds~\cite{Maggiore00}), 
${\cal J}^{(0)}(f) \simeq [2 S_h(f)/5 S_n(f)]^2$. 
Using the two approximate expressions for ${\cal J}^{(0)}(f)$, and
assuming that the main contribution to the 
SNR comes from a frequency band $\Delta f \sim f$, over which $S_n(f)$ and $S_h(f)$
are roughly constant, Eq.~(\ref{snrm1}) can be cast, respectively, in the form
\be
\SNR \approx 4.2\times 10^2\,\Theta_m^{(0)}\,
\left(\frac{\eta}{10^5}\right)^{1/2}
\label{snrm1gg}
\ee
and
\be
\SNR \approx 1.3\times 10^2\,\Theta_m^{(0)}\,
\left[\frac{\hc(f)}{\hrms(f)}\right]^2
\,\left(\frac{\eta}{10^5}\right)^{1/2}
\,,
\label{snrm1ll}
\ee
where $\eta \equiv \To\,\Delta f$;
as reference values we consider $\Delta f = 10^{-3}$ Hz and $\To = 10^{8}$ sec.
$\hc$ and $\hrms$ are the characteristic amplitude of the signal and the noise, respectively,
per logarithmic frequency interval (see {\it e.g.}~\cite{UV01,AR99}).
Notice that when $\hc \ll h_{\rm rms}$, the SNR is reduced by the factor 
$[\hc(f)/\hrms(f)]^2
\ll 1$ with respect to the case where the (isotropic component of the) signal dominates
the noise. Eq.~(\ref{snrm1gg}) represents the sensitivity limit of LISA to anisotropic backgrounds:
when $\hc (f) \simgt \hrms (f)$, the reduction of the instrumental noise does not improve the SNR.

In the LISA frequency band we expect a strong 
stochastic background generated by the galactic population 
of close binary systems, mainly white dwarf (WD) binaries~\cite{HBW90,BH97,PP98}. 
WD binary systems are mainly located in the galactic disk, and
the background that they generate is anisotropic for an observer in the solar system.
In order to gain insight into the LISA ability of detecting such signature
we model the WD distribution as being uniform within a cylinder of radius $r_0$ and height 
$2z_0$ (with $r_0\,>\,z_0$); typical values are $r_0\approx 5.5$ kpc and
$z_0 \approx 0.3$ kpc~\cite{HBW90}. By making this
approximation the computation of $P_{lm}$ and $\Theta_m^{(0)}$ simplifies considerably. 
For $S_h(f)$ and $S_n(f)$ we use the estimate provided in~\cite{HBW90,BH97} and~\cite{LISA_ppa},
respectively, whose analytical approximations are given by Eq.~(3.1) and (2.6) of~\cite{UV01}.
Eq.~(\ref{snrm1gg}) yields:
\be
\left(\frac{{\rm S}}{{\rm N}}\right)_{1} \approx 20\, \left(\frac{\eta}{10^5}\right)^{1/2}
\,,\,
\left(\frac{{\rm S}}{{\rm N}}\right)_{2} \approx 60\, \left(\frac{\eta}{10^5}\right)^{1/2}\,.
\label{snr1disk}
\ee
This result clearly shows that LISA can unambiguously identify the anisotropic component of
the stochastic background generated by galactic WD binaries.
By exploiting this signature one could also disentangle the galactic contribution 
from the extra-galactic one, which is isotropic to a large extent, being dominated by
sources at $z \sim 1$~\cite{SFMP00}. The large SNR at which detection can be made will 
enable us to extract valuable information about the distribution of binary systems
in the galaxy. However only the harmonics with $m=1$ and $2$ will clearly stand above the 
noise; for $m > 2$ the SNR is negligible. We have checked that the former results
are weakly sensitive on the approximations that we have made: Eq.~(\ref{snr1disk})  
agrees within $\approx 15\%$ with the value
obtained by the numerical integration of Eq.~(\ref{snrm1}) 
using Eq.~(\ref{J0}) over the 
entire sensitivity window and considering a more sophisticated source distribution. 
The galactic background from WD-WD binaries satisfies the condition $\hc(f) \gg \hrms(f)$ 
over the sensitivity window where most of the SNR is accumulated. If other 
galactic populations in the disk generate a stochastic background buried into the noise, 
LISA could in principle detect the signal, but at a SNR smaller than~(\ref{snr1disk}) by the
factor $\sim [\hc(f)/\hrms(f)]^2$, which makes rather unlikely to discriminate the two populations. 

LISA is very effective in detecting the anisotropy of
signals from sources located in the disk; one might wonder whether
populations characterized by a spherical distribution (such as MACHO binary systems in the Galactic
halo~\cite{ITN99}) would also produce a detectable anisotropy. We assume for sake of
simplicity that the sources are uniformly distributed within a sphere of radius $r_c$. 
Depending on the value of $r_c$, whether it is smaller or greater than $r_{\rm GC} \simeq 8.5$ kpc, 
the distance of the solar system from the Galactic Centre, the sources produce
a different degree of anisotropy. We consider first the case of a {\it bulge} distribution,
where $r_c < r_{\rm GC}$. Assuming $\hc(f) \gg \hrms (f)$, and $r_c = 1$ kpc,
Eq.~(\ref{snrm1gg}) yields:
\be
\left(\frac{{\rm S}}{{\rm N}}\right)_{1} \approx 132\, \left(\frac{\eta}{10^5}\right)^{1/2}
\,,\,
\left(\frac{{\rm S}}{{\rm N}}\right)_{2} \approx 13\, \left(\frac{\eta}{10^5}\right)^{1/2}\,.
\label{snr1b}
\ee
We have checked that if $r_c = 0.5$ kpc, the SNR is essentially
unaffected, and for $r_c = 7$ kpc the SNR becomes $\approx 78$ and 7 for $m=1$ and
2, respectively. If $\hc \simlt \hrms$ the values in Eq.~(\ref{snr1b}) are reduced by 
the factor $\approx 0.3\, [\hc(f)/\hrms(f)]^2$. For sources located
in an extended {\it halo}, with $r_c > r_{\rm GC}$, the multipole moments scale as 
$P_{lm} \sim (r_{\rm GC}/r_c)^l$, and Eq.~(\ref{snrm1gg}) yields:
\ba
\left(\frac{{\rm S}}{{\rm N}}\right)_{1} & \approx & 17 \, 
\left(\frac{8.5\,{\rm kpc}}{r_c}\right)^2\,\left(\frac{\eta}{10^5}\right)^{1/2}
\,,\nonumber\\
\left(\frac{{\rm S}}{{\rm N}}\right)_{2} & \approx & 10\, 
\left(\frac{8.5\,{\rm kpc}}{r_c}\right)^2\,\left(\frac{\eta}{10^5}\right)^{1/2}\,.
\label{snr1h}
\ea
LISA can therefore provide information on a background generated by sources characterized by a spherical
distribution. Eqs.~(\ref{snr1disk}), (\ref{snr1b}) and~(\ref{snr1h}) show that LISA can indeed
offer a radically new mean to explore the distribution of compact objects in the
galaxy. However, the task of pin-pointing the exact structure of the population will be highly
non-trivial~\cite{Cornish01m}. 

Although the former results are encouraging, they also suggest that it is unlikely to be able to gain
significant information on the much smaller degree of anisotropy produced 
by extra-galactic generated backgrounds, for which $\Theta_m^{(0)} \ll 1$, even if $h_c \gg h_{\rm rms}$,
see Eq.~(\ref{snrm1gg}). {\it A fortiori} the study of the anisotropy of the primordial GW
background is essentially ruled out. In principle one could attempt to detect
the dipole anisotropy induced by the motion of our local frame with $v/c\sim 10^{-3}$ with respect
to the Hubble flow. However LISA is not sensitive to $l=1$, and one
would have to rely on the next ``visible'' multipole moment ($l=2$). 
For $l=2$ we have $\Theta_m^{(0)} \simlt 10^{-6}$, and it 
would require a ridiculous time of integration to achieve SNR $> 1$.

The tremendous sensitivity ensured by 
the space technology, and the clear advantage of flying two independent LISA 
instruments~\cite{UV01} -- possibly with the option of achieving optimal sensitivity in
the frequency band 0.1 Hz $\simlt f \simlt$ 1 Hz, which is likely free from astrophysically 
generated backgrounds -- is stimulating further work regarding the new range of 
opportunities provided by cross-correlation experiments~\cite{Cornish01m,Cornish01s}.
The main advantage of this configuration for the study of 
the background anisotropy is the full sensitivity to the entire set of 
multipole moments $l$:
in general $\gamma_{lm}(f) \ne 0$ $\forall l$~\cite{AO97}. The value and frequency dependence 
of $\gamma_{lm}(f)$ depend on the actual separation and relative orientation of the two 
instruments, and one could
envisage to ``tune'' the configuration to some particular anisotropy structure. 
For two identical interferometers located at a distance $D$, the SNR reads 
(cfr. also~\cite{AO97})
\be
\SNR=\sqrt{\To}\,
\left\{\int_{0}^{\infty}df\,{\cal J}(f) \Theta_m^2(f) \right\}^{1/2}\,,
\label{snrm2}
\ee
where ${\cal J}(f)$ and $\Theta_m(f)$ are the analogous of ${\cal J}^{(0)}(f)$ and $\Theta_m^{(0)}$,
obtained by replacing $\gamma_0$ and $\gamma_{lm}^{(0)}$ with $\gamma(f)$ and 
$\gamma_{lm}(f)$, respectively. 

We have estimated the SNR that one could achieve by cross-correlating the data sets of  
two identical LISAs, placed at a distance $D=1$ AU, and for a few ``random'' orientations of the
instruments. The key difference with respect to the LISA mission is that the odd multipole 
moments, in particular $l=1$ and $l=3$, become visible; as a consequence the harmonics with $m=1$ 
will become stronger, and also the harmonic with $m=3$ could
be observable. As an example, for a disk distribution, one would achieve $({\rm S/N})_3  \simgt 3$;
for bulge and extended halo distributions $({\rm S/N})_1$ would be a factor $\approx 2$ and 10 higher 
than~(\ref{snr1b}) and~(\ref{snr1h}), respectively. 
This will impact on the ability of the instrument of gaining
insight on the source distributions, and disentangling the contributions from the disk, the bulge
and the halo. However this particular aspects deserve further investigation.

Cross-correlations would also make marginally possible to set limits on the anisotropy of the 
GW primordial background, thanks to the sensitivity to the dipole moment. 
Let us consider a primordial background with $P(\hat{\Omega})=1+\beta_D\hat{\Omega}\cdot\hat{\Omega}_D
+O(\beta_D^2)$, where the parameter $\beta_D$ measures the 
degree of anisotropy and the unit vector $\hat{\Omega}_D$ defines the 
direction of maximum anisotropy. Assuming that the primordial spectrum 
overwhelms both the intrinsic noise and the astrophysical background 
(i.e. $\hOgw \simgt 10^{-10}$) the parameter $\beta_D$ must satisfy 
$\beta_D > 8\times 10^{-3}$ $(\To/10^8\,\mbox{sec})^{-1/2}$ 
$(\Delta f/10^{-3}\,\mbox{Hz})^{-1/2}$
in order to detect the dipole component. The dipole anisotropy induced by the motion of the local
reference frame with respect to the Hubble flow corresponds to $\beta_D \sim 10^{-3}$; therefore
only an instrument with optimal sensitivity in the band $f \simgt 0.1$ Hz would be able to pick it up. 
This band presents also the considerable advantage of being completely transparent to primordial 
GW backgrounds~\cite{UV01}. Assuming that one can perfectly remove the induced dipole anisotropy from 
the data using independent information  provided by CMB experiments,  
one could then be able to test an {\it intrinsic} statistical departure from isotropy compatible with the measurement of the quadrupole anisotropy of the CMB temperature~\cite{cmb}.

\end{document}